\documentclass[12pt,preprint]{aastex}

\shorttitle{HCN in APM08279+5255}
\shortauthors{WAGG ET AL.}
\slugcomment{Accepted by ApJL}

\def\etals{{ et al. }\rm}

\begin{document}

\title{HCN J=5-4 Emission in APM08279+5255 at z=3.91 
\footnote{Based on observations carried out with the IRAM Plateau de
 Bure Interferometer. IRAM is supported by INSU/CNRS (France), MPG (Germany) 
and IGN (Spain).}}

\author{
J. Wagg,$^{1,2}$ D.J. Wilner,$^{1}$ 
R. Neri,$^{3}$ D. Downes,$^{3}$ and T. Wiklind$^{4}$ 
}

\affil{$^{1}$Harvard-Smithsonian Center for Astrophysics, 
             Cambridge, MA, 02138}
\email{jwagg@cfa.harvard.edu}

\affil{$^{2}$Instituto Nacional de Astrof\'isica, \'Optica y Electr\'onica 
             (INAOE), Aptdo. Postal 51 y 216, Puebla, Mexico}

\affil{$^{3}$Institut de Radio Astronomie Millim\'etrique, 
            St. Martin d'H\`eres, F-38406, France}

\affil{$^{4}$ESA Space Telescope Division, STScI, 
             3700 San Martin Drive, Baltimore, MD 21218, USA}

\begin{abstract}
We detect HCN J=5-4 emission from the ultraluminous quasar APM08279+5255 
at z=3.911 using the IRAM Plateau de Bure interferometer.
This object is strongly gravitationally lensed, yet still thought to be 
one of the most intrinsically luminous objects in the Universe. 
The new data imply a line luminosity 
$L'_{HCN(J=5-4)} = 4.0 \pm 0.5 \times 10^{10}$~K~km~s$^{-1}$~pc$^2$.
The $\sim440$~km~s$^{-1}$ full width half maximum of the HCN J=5-4 line 
matches that 
of the previously observed high-J CO lines in this object and 
suggests that the emission 
from both species emerges from the same region: 
a warm, dense circumnuclear disk. 
Simple radiative transfer models suggest an enhanced abundance of
HCN relative to CO in the nuclear region of APM08279+5255, perhaps due to 
increased ionization, or possibly the selective depletion of oxygen.
The ratio of far-infrared luminosity to HCN luminosity is at the
high end of the range found for nearby star forming galaxies, but
comparable to that observed in the few high redshift objects detected 
in the HCN J=1-0 line.  This is the first clear detection of high-J 
HCN emission redshifted into the 3-millimeter atmospheric window. 
\end{abstract}


\section{Introduction}

The quasar APM08279+5255, serendipitously discovered by Irwin et al.\ (1998), 
has become an archetypical ultraluminous high redshift source for study 
at all wavelengths, from X-ray to radio (e.g. Ellison et al.\ 1999,
Gallagher et al.\ 2002, Soifer et al.\ 2004).
The popularity of this source may be 
attributed, in part, to its tremendous apparent luminosity, of which
$\sim$10$^{15}$ L$_{\odot}$ emerges in the far-infrared (Lewis et al.\ 1998).
Optical imaging reveals multiple source components and suggests that the
 object is strongly gravitationally lensed 
(Ledoux et al.\ 1998, Ibata et al.\ 1999, Egami et al.\ 2000). 
Even after correction for the lensing amplification, 
APM08279+5255 remains one of the most intrinsically luminous objects known
 in the Universe. 

The extreme luminosity of APM08279+5255 is provided by a combination of 
an active galactic nucleus (AGN) and star formation activity.
Downes et al.\ (1999, hereafter D99) used the IRAM Plateau de Bure
interferometer (PdBI) to detect dust continuum and CO J=4-3 and J=9-8
emission at arcsecond scales in APM08279+5255, proposing that
 the emission originates in a warm, dense circumnuclear disk of sub-kpc size. 
APM08279+5255 is the only high redshift source where CO emission from
such highly excited levels as J=9 has been observed, indicative of 
extreme conditions in the molecular gas. In order to better understand the 
excitation of the molecular material and its physical nature, 
observations of additional diagnostic lines are needed. 

The HCN molecule, which requires substantially higher densities 
($n_{H_2} > 10^{4}$~cm$^{-3}$) than CO for collisional excitation due to 
its large dipole moment, 
is a promising tracer of environments such as the nucleus of APM08279+5255.
In the Galaxy, HCN emission selectively traces the high density gas 
where star formation takes place (Helfer \& Blitz 1997). 
In nearby star forming galaxies, the HCN J=1-0 line luminosity is 
found to correlate tightly with infrared/far-infrared luminosity 
(Solomon, Downes \& Radford 1992a, Gao \& Solomon 2004a, 2004b). 
At high redshifts, detection of HCN emission has proven challenging 
for the current generation of radio telescopes. The J=1-0 line has
been clearly detected in H1413+117 (the strongly lensed ``Cloverleaf'' 
at z=2.6, Solomon et al.\ 2003), and in the strongly lensed 
IRAS F10214+4724 at z=2.3 (Vanden Bout et al.\ 2004), and tentatively 
detected in J1409+5628 at z=2.6 (Carilli et al.\ 2005). 
The higher excitation HCN J=4-3 line was also tentatively detected in 
H1413+117 (Barvainis et al. 1997, although see Solomon \etals 2003).
Here we present the detection of strong HCN J=5-4 line emission in
APM08279+5255 at z=3.9 and discuss the implications for the physical 
conditions in this remarkable object.

We adopt the following $\Lambda$-dominated cosmological parameters:
 $H_0 = 70$ km s$^{-1}$ Mpc$^{-1}$, $\Omega_\Lambda = 0.7$, $\Omega_m = 0.3$
(Spergel et al.\ 2003).

\section{Observations}

The IRAM PdBI was used to observe APM08279+5255 in the HCN J=5-4 line, 
redshifted to the 3-millimeter band, on six dates in 1999 and 2001. 
All of the observations were made in compact configurations of the four or 
five available antennas. The phase center was offset by $1.25''$ from the 
CO peak position.
The receivers were tuned to 90.229 GHz, corresponding to
the HCN J=5-4 line (rest frequency of 443.1162 GHz) at z=3.911,
determined from the earlier CO line observations.
Spectral correlators covered a velocity range of $\sim$1600 km~s$^{-1}$. 
The baseline lengths ranged from 17 to 81 meters, leading to a 
beam size of $7.4''\times6.2''$ (position angle 79$^{o}$).
The nearby quasar 0749+540 was used for complex gain calibration. 
The flux scale was set using standard sources and other
strong sources including MWC349, CRL618, 3c345.3 and 0923+392,
and should be accurate to better than 20\%.
The integration time on source was about 32 hours (equivalent to 11.9 hours 
with the current six antenna array). The average rms noise is 
0.47~mJy~beam$^{-1}$ in each 100~km~s$^{-1}$ channel. 

\section{Results}

\begin{figure}
\centering
\includegraphics[height=6.0in,angle=270]{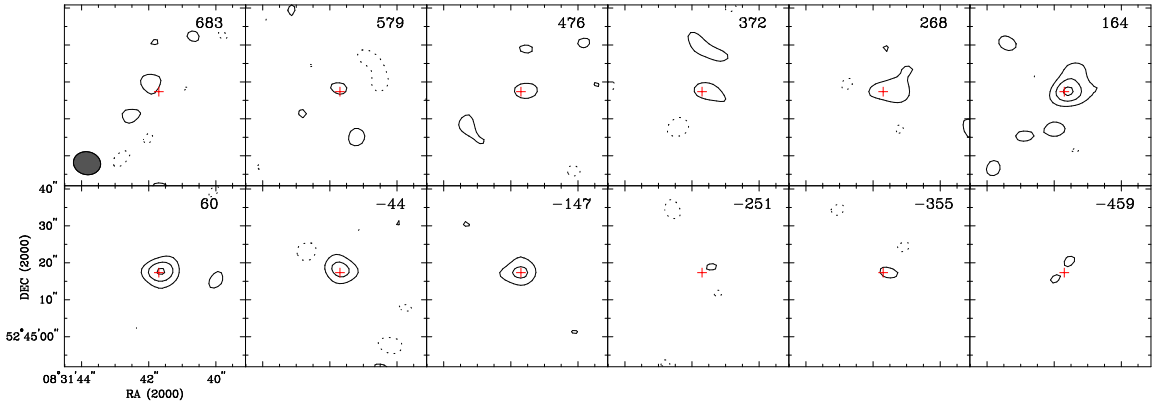}
\caption{
Channel maps of 90.2 GHz redshifted HCN J=5-4 emission plus 
continuum emission in APM08279+5255. The cross marks the position of the 
CO peak position from D99, offset by ($1\farcs19$, $-0\farcs36$) from the phase center 
(08$^h$31$^m$41$^s$.57, 52$^o$45$^m$17$^s$.7). 
The contour intervals are -2, 2, 4, and 6$\times\sigma$ 
(0.47 mJy~beam$^{-1}$). The filled ellipse shows the 
$7\farcs4 \times 6\farcs2$ PA 79.0$^{\circ}$ synthesized beam. 
The numbers in each panel indicate the velocity relative to z=3.911.
}
\label{fig:hcn_chan}
\end{figure}

\begin{figure}
\centering
\includegraphics[width=5in]{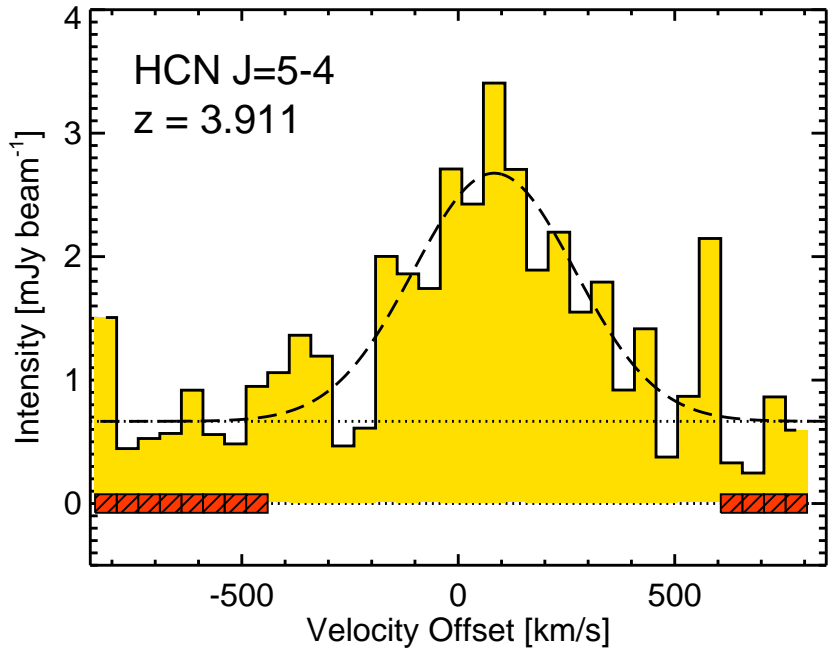}
\caption{
Spectrum of HCN J=5-4 emission from APM08279+5255 at 90.229 GHz, where
the velocity scale is relative to the CO redshift of 3.911.
The hatched regions mark the channels assumed to represent the 
line-free continuum level. The dashed line shows a Gaussian fit.
\label{fig:hcn_spec}}
\end{figure}

Figure~\ref{fig:hcn_chan} shows a series of 100~km~s$^{-1}$ channel maps 
across $\sim$900 km~s$^{-1}$ of the bandwidth.
Compact HCN J=5-4 line emission is clearly evident.
Figure~\ref{fig:hcn_spec} shows the spectrum at the position of peak 
intensity in Figure~\ref{fig:hcn_chan}. This position is consistent with that 
found by D99 for the CO emission within the uncertainties. 
The bandwidth is barely adequate to span the full extent of the HCN line
emission. As indicated in Figure~\ref{fig:hcn_spec}, we adopt narrow 
continuum regions at both ends of the spectrum to derive the HCN line 
properties. The 90.2 GHz continuum flux from the combination of these
``line-free'' spectral regions is 
$0.66\pm0.18$~mJy, in line with the value of 
$1.2\pm0.3$~mJy at 93.9 GHz found by D99, within the uncertainties.
A Gaussian fit to the continuum subtracted spectrum gives a 
peak flux of $2.01\pm0.51$ mJy~beam$^{-1}$,
central velocity $v_0 = 83 \pm 26$ km~s$^{-1}$
(z=$3.9124\pm0.0004$), and $\Delta V_{\rm FWHM} = 440 \pm 59$ km~s$^{-1}$.
The HCN J=5-4 line integrated intensity determined by integrating
over the velocity range $-200$ to $+500$~km~s$^{-1}$
is $0.98 \pm 0.12$ Jy~km~s$^{-1}$.  The line luminosity is 
$L'_{HCN(J=5-4)} = 4.0 \pm 0.5 \times 10^{10}$ K~km~s$^{-1}$~pc$^2$
using equation (3) of Solomon et al.\ (1992b). Table~1 lists the fitted and 
derived HCN line parameters.

\section{Discussion}

\subsection{HCN and CO emission region}

The HCN J=5-4 central velocity and line width, 
$\Delta V_{\rm FWHM} = 440 \pm 59$ km~s$^{-1}$, 
are very similar to that found for the CO J=4-3 line,
$\Delta V_{\rm FWHM} = 480 \pm 35$ km~s$^{-1}$ (D99).
This close correspondence suggests a co-spatial origin for 
the emission from these lines. Both species likely arise from 
the same dynamical structure, presumably a warm, dense 
circumnuclear disk.

If the high-J HCN and high-J CO lines emerge from the same region, then 
it is reasonable to assume that the HCN luminosity is magnified by 
gravitational lensing by the same factor as the CO luminosity.  
D99 argue that the magnification factor is 7 to 20 for an 
intrinsic CO source size of 160 to 270 pc 
(173 to 290 pc for the cosmology adopted here).  
Lewis et al.\ (2002) argue for a somewhat lower magnification factor 
of $\sim3$ based on the brightness of CO J=1-0 and CO J=2-1 lines
in sub-arcsecond resolution VLA images (Papadopoulos et al.\ 2001,
Lewis et al.\ 2002). Since the low-J CO line shapes are not well measured,
it is not clear if the low-J and high-J CO lines are co-spatial.
In any case, there is considerable uncertainty in the lensing model, 
and a wide range of magnification factors are viable.
Observations that spatially resolve the high-J CO line emission 
would be useful to directly probe the line brightness and distribution.

\subsection{HCN/CO ratio and Physical Conditions}
\label{sec:physical}
The $L'$ ratio for two lines equals the intrinsic brightness temperature
ratio averaged over the velocity and physical extent of the source,
and may be used to constrain the 
physical conditions of the gas (Solomon, Downes \& Radford 1992b).
D99 reproduce the CO J=9-8/J=4-3 line luminosity ratio of $0.48\pm0.07$ 
using escape-probability radiative transfer and deduce a gas temperature, 
T$_{\rm kin}$ $\sim$140 - 250 K and 
a molecular hydrogen density, n$_{\rm H_2} \sim4,000$~cm$^{-3}$. 
The CO J=9-8/J=4-3 ratio and the CO J=4-3/J=1-0 ratio also may be 
reproduced by gas with a lower temperature and higher density.  
Using the \textit{RADEX}\footnote{
http://www.strw.leidenuniv.nl/\~{}moldata/radex.html} 
(Sch\"{o}ier et al.\ 2005) software, we find that 
T$_{\rm kin}$ $\sim$80~K and n$_{\rm H_2}$ $\sim40,000$~cm$^{-3}$
matches the CO line ratios, with 
$N(CO)/\Delta v = 4\times10^{17}$~cm$^{-2}$~(km~s$^{-1}$)$^{-1}$. 
This single component model also provides a good fit to all the 
observed lines in the CO ladder, up to the J=11-10 transition, 
as shown by Wei{\ss} et al.\ (2005, in prep.) who reach similar
conclusions about the gas density and temperature.

These radiative transfer calculations assume that the HCN molecules are
excited by collisions. It is possible that radiative excitation plays a
role, as infrared radiation from dust heated by the AGN may excite HCN
through stretching and bending vibrational modes at (rest frame)
3, 5 and 14~$\mu$m. However, as argued by D99 for CO excitation in this
source, the large dust mass and consequent high opacity likely prohibits
the short wavelength infrared radiation from affecting a large volume
of the gas.

If we assume the conditions inferred from the CO line ratios,
and a nominal [HCN/CO] abundance ratio of $10^{-3}$, as found 
in Galactic star forming cores (Helfer \& Blitz 1997),
then the single component radiative transfer model predicts a magnified 
luminosity $L'_{HCN(J=5-4)} = 3.9 \times 10^{9}$ K~km~s$^{-1}$~pc$^2$,
which is a factor of 10 lower than the observed value.
A plausible way to produce a higher HCN J=5-4 line luminosity is 
with a higher [HCN/CO] abundance ratio. From the radiative transfer 
model, this ratio must be in the range (1 - 2) $\times 10^{-2}$ 
to match the data.
There are at least two plausible scenarios that could lead to a 
higher [HCN/CO] abundance ratio:
(1) an increased ionization rate in the vicinity of the AGN that
enhances the HCN abundance (Lepp \& Dalgarno 1996); 
(2) selective depletion of oxygen relative to carbon in the 
nuclear region.  A combination of these two effects has been 
proposed to explain observations of HCN and CO in the nearby 
active nucleus NGC~1068 
(Sternberg et al.\ 1994, Shalabiea \& Greenberg 1996, Usero et al.\ 2004).
If there has been selective depletion of oxygen in APM08279+5255, 
then the super-solar Fe-to-O ratio found from X-ray data 
(Hasinger et al.\ 2002) may not provide a relevant constraint 
on metal enrichment (or age) at high redshift. 

Observations of additional HCN transitions would allow the 
physical conditions of the HCN emitting material to be 
determined independently of the constraints provided by CO observations.
The single component model (with enhanced [HCN/CO] = $10^{-2}$) 
predicts $L'_{HCN(J=1-0)} = 4.4 \times 10^{10}$ K~km~s$^{-1}$~pc$^2$.
A secure detection of the HCN J=1-0 line 
at this luminosity (redshifted to 18.05 GHz) would require $\sim100$ hours 
of integration time using a 100 meter diameter class radio telescope. 

The mass of dense gas may be estimated from the HCN line luminosity, 
$M_{dense} = \alpha L'_{HCN(J=1-0)}$, where $\alpha$ for warm gas 
($\sim50$~K) is $\sim7$~M$_{\odot}$(K~km~s$^{-1}$~pc$^2)^{-1}$ 
(Gao \& Solomon 2004b). For the HCN J=1-0 line luminosity estimated
from our radiative transfer calculation,
$M_{dense} = 3.1\pm0.4 \times 10^{11}$~M$_{\odot}$
uncorrected for amplification by gravitational lensing. 
A conventional estimate of the (total) molecular gas mass 
from the nuclear CO line luminosity (Lewis et al. 2002)
results in a comparable value, i.e.
$M_{H_2} = \beta L'_{CO(1-0)}$, where
$\beta \sim1-5$~M$_{\odot}$(K~km~s$^{-1}$~pc$^2)^{-1}$ 
(Downes \& Solomon 1998, Young \& Scoville 1991).
In this regard, APM08279+5255 is somewhat similar 
to IRAS~F10214+4724, where the molecular medium is also dense.

\subsection{Far-infrared, CO and HCN Luminosities}

According to Solomon et al.\ (2003), the presence of strong HCN emission 
from a high redshift object is an indicator of an ongoing starburst.
A tight correlation exists between far-infrared luminosity and HCN J=1-0 
line luminosity in star forming galaxies (Solomon, Downes \& Radford 1992).  
The far-infrared luminosity is primarily due to dust heated by young, 
high mass stars. To the extent that AGN activity is not important, 
the ratio $L_{FIR}/L'_{CO}$ measures the star formation rate per unit 
molecular gas mass, while the ratio $L_{FIR}/L'_{HCN}$ measures the star 
formation rate per unit dense molecular gas. For the sample of star forming 
galaxies examined by Gao \& Solomon (2004a, 2004b),
$L_{FIR}/L'_{CO}$ increases with far-infrared luminosity, 
while $L_{FIR}/L'_{HCN}$ does not.

To examine APM08279+5255 in the context of the correlations established 
from nearby galaxies, we must appeal to the results of the radiative 
transfer calculation (\S~\ref{sec:physical}) to predict the HCN J=1-0 line 
luminosity, since this line was not observed.
For galaxies in the local Universe, generally only the 
HCN J=1-0 line has been observed; the higher frequency HCN J=5-4 line 
is not accessible to ground based observations due to a strong water 
absorption feature in the Earth's atmosphere.

Observed values of $L_{FIR}/L'_{HCN}$ for star forming galaxies are 
typically less than 2000 L$_{\odot}$~(K~km~s$^{-1}$~pc$^2)^{-1}$.
The high redshift objects with secure HCN J=1-0 detections are
H1413+117, F10214+4724 and J1409+5628, and show $L_{FIR}/L'_{HCN}$
of 1700, 3000 and 5000, respectively (Carilli et al.\ 2005).
For APM08279+5255, the spectrum has been decomposed into 
starburst and AGN components by Rowan-Robinson (2000) 
who concludes $L_{FIR} = 9.9\times 10^{13}$~L$_{\odot}$ (under our assumed
 cosmology) from the starburst. Adopting this value, 
$L_{FIR}/L'_{HCN} = 2300$, similar to the ratios observed in the other 
high redshift objects and at the high end of the ratios observed in local 
galaxies. This ratio is sensitive to the detailed decomposition of the 
spectrum and would be higher if a higher fraction of $L_{FIR}$ 
were due to star formation rather than AGN activity.
This calculation also does not account for the possible effects of 
differential lensing, and it is plausible that differential lensing 
results in a higher magnification factor for the far-infrared emission 
than for the HCN emission, since the molecular gas may have a larger 
spatial extent (D99), and that could lead to an enhanced apparent value. 

Perhaps the nearest analog to APM08279+5255 in the local 
Universe is Mrk231, an ultraluminous infrared galaxy (ULIRG) with an 
AGN, whose bolometric luminosity is dominated by a starburst 
(Solomon, Downes \& Radford 1992a, Davies, Tacconi \& Genzel 2004). 
In Mrk231, $L'_{HCN}/L'_{CO} \sim 0.25$, the highest in the 
Gao \& Solomon (2004a) sample.
In APM08279+5255, $L'_{HCN}/L'_{CO} = 0.34\pm0.09$, using the
(inferred) HCN J=1-0 luminosity and the (observed) nuclear 
CO J=1-0 luminosity (Papadopoulos et al.\ 2001). This is at   
the high end of the range 0.10 - 0.25 found for local galaxies
(Gao \& Solomon 2004a, 2004b). 
The fact that both Mrk231 and APM08279+5255 exhibit high 
$L'_{HCN}/L'_{CO}$ values suggests a possible connection between 
the presence of an AGN and a high dense gas mass fraction.
A physical mechanism to explain this connection might be that
radiation pressure leads to an increased gas density in the vicinity 
of the AGN, as has been suggested for H1413+117 (Barvainis et al.\ 1997).

\section{Summary}

We have detected HCN J=5-4 emission from the ultraluminous 
quasar APM08279+5255 at z=3.91.
The similarity of the HCN J=5-4 line profile to the CO J=4-3 and
J=9-8 lines suggests the emission from these species is co-spatial. 
We have used a simple radiative transfer model to constrain the 
physical conditions in the molecular gas.  
The warm (T$_{kin} \sim$80~K), dense (n$_{H_2} \sim 40,000$~cm$^{-3}$) 
gas is presumably located in a circumnuclear disk.
However, the [HCN/CO] abundance ratio appears to be enhanced. 

This is the first clear detection of high-J HCN emission redshifted 
into the 3-millimeter atmospheric window. Such observations of high-J
transitions of dense gas tracers are expected to become routine with 
the Atacama Large Millimeter Array, which will provide an improvement 
in sensitivity by more than an order of magnitude. Observations of dense gas 
tracers in large samples of high redshift ultraluminous galaxies will help to
 probe their power sources, and star formation/AGN activity.

\section{Acknowledgments}

We thank the IRAM PdBI staff for carrying out these observations. 
J.W. thanks the SAO for support through a predoctoral student fellowship and 
the Department of Astrophysics at INAOE for a graduate student scholarship.
This work was partially supported by CONACYT grant 39953-F.
We thank Floris van der Tak for helpful discussions about radiative transfer
and the \textit{RADEX} team for making their software available. 
We are grateful to the anonymous referee for promptly providing helpful 
suggestions and comments.

\clearpage

\begin{deluxetable}{lc}
\tablecaption{HCN J=5-4 line parameters for APM08279+5255.
\label{table1}}
\tablewidth{0pt}
\tablehead{\colhead{} & \colhead{}}
\startdata
HCN J=5-4 peak:        & $2.01\pm0.51$~mJy \\
HCN J=5-4 $\Delta V_{\rm FWHM}$: &  $440\pm59$~km~s$^{-1}$  \\
HCN J=5-4 $v_0$\tablenotemark{a}: & $83\pm26$~km~s$^{-1}$ \\
$I_{HCN(J=5-4)}$: &  $0.98\pm0.12$~Jy~km~s$^{-1}$  \\
$L'_{HCN(J=5-4)}$\tablenotemark{b}: & $4.0\pm0.5 \times 10^{10}$~K~km~s$^{-1}$~pc$^2$  \\
90.2 GHz continuum:    & $0.66\pm0.18$~mJy \\
\enddata            
\tablenotetext{a}{Velocity with respect to z=3.911, determined from the CO 
lines.}
\tablenotetext{b}{The HCN J=5-4 luminosity is uncorrected for lensing
 magnification.}
\end{deluxetable}

\clearpage

\clearpage

\end{document}